\documentclass[%
 preprint,
 amsmath,amssymb,
 aps, physrev,
showkeys,
]{revtex4-2}

\usepackage{graphicx}% Include figure files
\usepackage{dcolumn}% Align table columns on decimal point
\usepackage{bm}% bold math
\begin{document}

\newcommand{\arcsec}{\ifmmode^{\prime\prime}\else $^{\prime\prime}$\fi}
\newcommand{\arcmin}{\ifmmode^{\prime}\else $^{\prime}$\fi}
\newcommand{\degrees}{\ifmmode^{\circ}\else $^{\circ}$\fi}
\newcommand{\kms}{km\,s$^{-1}$}
\newcommand{\ms}{m\,s$^{-1}$}
\newcommand{\micron}{$\mu$m}
\newcommand{\smallhalf}{\textstyle \frac{1}{2} \displaystyle}
\newcommand{\smallonethird}{\textstyle \frac{1}{3} \displaystyle}
\newcommand{\smallonefourth}{\textstyle \frac{1}{4} \displaystyle}
\newcommand{\smalltwothirds}{\textstyle \frac{2}{3} \displaystyle}
\newcommand{\smallfourthirds}{\textstyle \frac{4}{3} \displaystyle}
\newcommand{\smallonequarter}{\textstyle \frac{1}{4} \displaystyle}
\newcommand{\bfx}{{\bf x}}
\newcommand{\bfk}{{\bf k}}
\newcommand{\bfK}{{\bf K}}
\newcommand{\bfc}{{\bf c}}
\newcommand{\bfe}{{\bf e}}
\newcommand{\bff}{{\bf f}}
\newcommand{\bfg}{{\bf g}}
\newcommand{\bfq}{{\bf q}}
\newcommand{\bfr}{{\bf r}}
\newcommand{\bfu}{{\bf u}}
\newcommand{\bfU}{{\bf U}}
\newcommand{\bfv}{{\bf v}}
\newcommand{\bfw}{{\bf w}}
\newcommand{\bfb}{{\bf b}}
\newcommand{\bfB}{{\bf B}}
\newcommand{\bfH}{{\bf H}}
\newcommand{\bfI}{{\bf I}}
\newcommand{\bfE}{{\bf E}}
\newcommand{\bfD}{{\bf D}}
\newcommand{\bfj}{{\bf j}}
\newcommand{\bfa}{{\bf a}}
\newcommand{\bfn}{{\bf n}}
\newcommand{\bfs}{{\bf s}}
\newcommand{\bfA}{{\bf A}}
\newcommand{\bfF}{{\bf F}}
\newcommand{\bfM}{{\bf M}}
\newcommand{\bfP}{{\bf P}}
\newcommand{\bfQ}{{\bf Q}}
\newcommand{\bfS}{{\bf S}}
\newcommand{\bfT}{{\bf T}}

\newcommand{\bfgamma}{{\mbox{\boldmath $\gamma$}}}
\newcommand{\bfkappa}{{\mbox{\boldmath $\kappa$}}}
\newcommand{\bfOmega}{{\mbox{\boldmath $\Omega$}}}
\newcommand{\bfomega}{{\mbox{\boldmath $\omega$}}}
\newcommand{\bfsigma}{{\mbox{\boldmath $\sigma$}}}
\newcommand{\bftau}{{\mbox{\boldmath $\tau$}}}
\newcommand{\bfxi}{{\mbox{\boldmath $\xi$}}}

\newcommand{\bfnu}{{\mbox{\boldmath $\nu$}}}

\newcommand{\bfnabla}{{\mbox{\boldmath $\nabla$}}}

\newcommand{\bfitl}{{\mbox{\boldmath $l$}}}

\newcommand{\calC}{{\mathcal C}}
\newcommand{\calS}{{\mathcal S}}
\newcommand{\calV}{{\mathcal V}}
\newcommand{\calE}{{\mathcal E}}
\newcommand{\calP}{{\mathcal P}}
\newcommand{\calR}{{\mathcal R}}
\newcommand{\calT}{{\mathcal T}}
\newcommand{\calH}{{\mathcal H}}
\newcommand{\calO}{{\mathcal O}}

%\preprint{APS/123-QED}

\title{\textbf{Volumetric effects in viscous flows in circular and annular tubes with wavy walls}}

\author{Yisen Guo}

\author{John H. Thomas}
\email{Contact author: thomas@me.rochester.edu}

\affiliation{Department of Mechanical Engineering, University of Rochester, Rochester, NY 14627, USA}

%\date{\today}

\begin{abstract}
We point out that, in the usual way of specifying a sinusoidal waviness of the wall of a tube of circular cross section, in which the mean radius is kept constant, the interior volume of the tube increases with increasing wave amplitude. We compare this case with the case where the interior volume is kept constant by reducing the mean radius as the wave amplitude increases. We present and compare numerical results of these two cases for steady, pressure driven, laminar viscous flow in a tube with a stationary wavy wall, for both circular and annular tubes. The volume flow rate and the hydraulic resistance can differ in the two cases by as much as 10\% for wave amplitudes as small as 20\% of the mean radius and as much as 50\% for larger wave amplitudes. For a circular tube, we derive a scaling law that relates the two cases based on dimensional analysis, allowing the behavior in the constant-volume case to be determined from that in the constant-mean-radius case. Additionally, we consider peristaltic pumping due to a moving sinusoidal wall wave and show that the volume-change effect is significant even at small wave amplitudes, and that the volume flow rates in the two cases can differ significantly, by as much as 50\% as the wave amplitude approaches its maximum value.
\end{abstract}

%\keywords{Suggested keywords}%Use showkeys class option if keyword %display desired

\maketitle

\section{Introduction}
A problem of broad interest in the realm of incompressible viscous flow is to determine the effect of a wavy displacement of the wall of a duct or tube on a flow within it. This problem arises in a few different contexts. One such context is to determine the velocity field, volume flow rate, and hydraulic resistance for a steady flow driven by an axial pressure gradient in a circular tube with radius varying sinusoidally along its axis \cite{belinfante1962, dodson1971, chow1972, lessen1976, deiber1979, ralph1987, hemmat1995, sisavath2001, wang2006}. Another such context is to determine the flow field and mean flow rate driven by peristalsis in the case of a sinusoidal wall wave propagating in the axial direction \cite{burns1967, shapiro1969,jaffrin1971,takabatake1988,wang2011,carr2021}. In the case of a channel or duct bounded by plane walls, sinusoidal displacements of any of the walls, without changing their mean position, leaves the interior volume over an integral number of wavelengths unchanged. However, in the case of an open or annular tube of circular cross-section, adding a sinusoidal displacement of the wall, without changing the mean radius, changes the interior volume of the tube. This simple geometrical fact has been generally overlooked in studies of these flows.

%\section{The circular tube}

\begin{figure}
  \centerline{\includegraphics[width=0.6\textwidth]{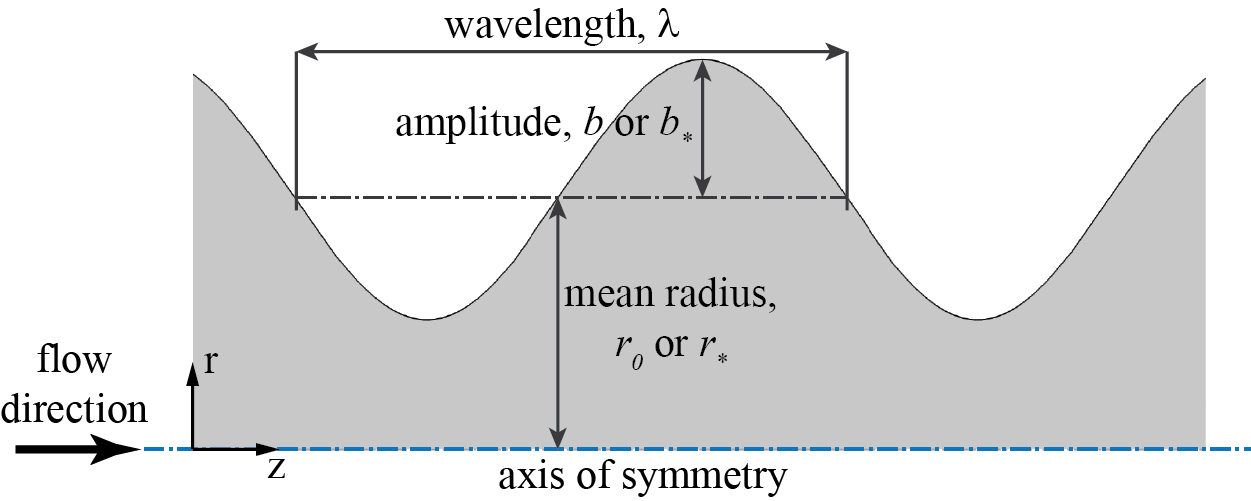}}
  \caption{Schematic diagram of an circular tube with a wavy wall (cf. equation \ref{eq:sinusoid}).
  }
 \label{fig:schematic}
\end{figure}

Consider a tube of circular cross section with a sinusoidally varying radius of the wall, as shown in Fig.\ \ref{fig:schematic}. The radius $r(z)$ of the tube wall is given by the usual expression
	\begin{equation}
	r(z) = r_0 + b\sin(2\pi z/\lambda) ,
	\label{eq:sinusoid}
	\end{equation}
where $r_0$ is the fixed mean radius and $b$ and $\lambda$ are the amplitude and wavelength of the sinusoidal displacement. The axially varying cross-sectional area $A(z)$ is given by 
	\begin{equation}
	A(z) = \pi [r_0 + b\sin(2\pi z/\lambda)]^2 = \pi [r_0^2 + 2 r_0 b \sin(2\pi z/\lambda) 
	+ b^2 \sin^2(2\pi z/\lambda)] .
	\end{equation}
The total internal volume $V_\lambda$ over one wavelength along the tube is then given by 
	\begin{equation} 
	V_\lambda = \int_0^\lambda A(z) dz = \pi [ r_0^2 + \frac{b^2}{2} ] \lambda ,
    \label{eq:volume1}
	\end{equation} 
and we see that the addition of the sinusoidal displacement increases the interior volume of the tube by an amount $\pi b^2 \lambda/2$ over each wavelength. In other words, the average cross-sectional area over one wavelength of the tube is increased from $\pi r_0^2$ to $\pi (r_0^2 + b^2/2)$ by the wavy displacement. Note that adding azimuthal dependence of the form $r(z) = r_0 + b\sin(2m\theta)\sin(2\pi z/\lambda)$, where $m$ is an integer, in order to produce a bumpy wall \cite{wang2006}, produces no additional change in volume beyond that given by equation (\ref{eq:volume1}).
%Imagine a thin, axisymmetric ring of volume within the undeformed tube. After In places where the tube is expanded ($0 < 2 \pi z/\lambda < \pi$), the radius of the ring, and hence its circumference, is increased, while in places where the cross-section is contracted ($\pi < 2 \pi z/\lambda < 2\pi$), the circumference is decreased, so the volume increase in the expanded section exceeds the volume decrease in the contracted section, thus producing a net volume increase \footnote{Note that adding azimuthal dependence of the form $r(z) = r_0 + b\sin(2m\theta)\sin(2\pi z/\lambda)$, where $m$ is an integer, in order to produce a bumpy wall \cite{wang2006}, produces no additional change in volume beyond that given by equation (\ref{eq:volume1}).}.

For a finite tube $n$ wavelengths long, the total volume will be $V = n V_\lambda$ and the volume added by the displacement will be $n \pi b^2 \lambda/2$. The added volume scales as $b^2$, and so, in some contexts, it might be negligibly small for very small displacements, but it is significant for larger displacements: for example, in the extreme case where $b = r_0$ and the tube is pinched off into a series of separate compartments, we have $V_\lambda = V_{\lambda {\rm max}} =3 \pi r_0^2 \lambda/2$, and hence the volume is increased by 50\%. 

Although there are many published papers on flows in tubes in which a sinusoidal wall displacement of the form (\ref{eq:sinusoid}) is imposed, we have found no mention of the associated volume change, let alone any analysis of its effects. Here we discuss some of the consequences of this volume change for the two types of flow mentioned above: steady flow driven by an axial pressure gradient in tubes with stationary wavy walls and peristaltic flow in tubes with a propagating wall wave. We consider both circular and annular tubes, with no curvature of their central axis, and we limit the discussion to incompressible, laminar, viscous flow. 

%The volume of fluid inside the tube increases as the amplitude $b$ of the displacement increases. 
%If the fluid is essentially incompressible, then, as the amplitude is increased, additional fluid must enter from an open end of the tube, or, if both ends of the tube are closed, then cavitation would occur. If the fluid is a gas, then in a closed tube the gas will expand to fill the increased volume and its density will decrease.

\section{The circular tube}

In this section we consider flows in a circular tube (i.e., an open tube of circular cross section), as shown in Fig.\ \ref{fig:schematic}, postponing discussion of annular tubes to the following section.

\subsection{Laminar viscous flow along the tube}
\noindent Consider the case of fully-developed (no entrance effects) laminar viscous flow of an incompressible fluid along the tube, driven by an axial pressure gradient, and suppose that we wish to determine how the velocity field, volume flow rate $Q$ and hydraulic resistance $\mathcal{R}$ (per unit length) depend on the amplitude and wavelength of the wavy wall. In the absence of the wavy displacement ($b=0$), the flow is simple Poiseuille flow, and we have
	\begin{equation}
	u_z(r) = \frac{1}{4\mu}\left( - \frac{\partial p}{\partial z}\right) (r_0^2 - r^2), \quad 
	Q = \frac{\pi r_0^4}{8\mu} \left( - \frac{\partial p}{\partial z}\right) , \quad
	\mathcal{R} \equiv \left( - \frac{\partial p}{\partial z}\right) \frac{1}{Q} = \frac{8\mu}{\pi r_0^4},
	\end{equation}
where $u_z(r)$ is the axial velocity, $p$ is the pressure, and $\mu$ is the dynamic viscosity. Adding a sinusoidal displacement of the form (\ref{eq:sinusoid}) to the wall of the tube will add a radial component of velocity, $u_r(r)$, produce pressure variations $p(r,z)$ and additional components of shear stress, increase the hydraulic resistance, and decrease the volume flow rate. Also, the axial pressure gradient $\partial p/\partial z$ will no longer be uniform, but instead will vary along the tube. 

The hydraulic resistance will increase monotonically with increasing amplitude $b$. From the fact that the hydraulic resistance scales inversely as the fourth power of the radius in a uniform tube, we can deduce that this increase in resistance is mostly due to the large increase in shear stress in the narrowed parts of the tube, which dominates the smaller decrease in shear stress in the widened parts of the tube. But the resistance will also be affected by the increase in total interior volume of the tube noted above. The increase in volume corresponds to an increase in the average cross-sectional area of the tube and hence an increase in the average distance of fluid particles from the wall of the tube, thereby producing a lower average shear stress compared to the constant-volume case. This reduction in shear stress will offset to some extent the increase in hydraulic resistance due to the constrictions produced by the sinusoidal displacement. 
%Argument above needs revising in light of Jessica's lubrication model -- cite our paper.

As an alternative to keeping the mean radius $r_0$ constant, we can keep the interior volume of the tube constant by decreasing the mean radius of the tube to a smaller value $r_*$ as the wave amplitude $b$ increases, that is, require 
%while scaling the amplitude of the disturbance to a smaller value $b_*$ in order to keep the relative amplitude the same. That is, require
	\begin{equation} 
	V_\lambda = \pi (r_*^2 + \frac{b^2}{2} ) \lambda = \pi r_0^2 \lambda, %\quad {\rm and} \quad \frac{b_*}{r_*} = \frac{b}{r_0}.
    \label{eq:correctedradius1}
	\end{equation}
or
\begin{equation}
	\frac{r_*}{r_0}= \sqrt{1 - \frac{1}{2} (\frac{b}{r_0})^2} \ < 1 \ .
	\label{eq:correctedradius2}
	\end{equation}
	%\begin{equation}
	%\frac{r_*}{r_0} = \sqrt{ \frac{1}{1 + \frac{1}{2} (\frac{b}{r_0})^2} } \ < 1 \ .
	%\label{eq:correctedradius}
	%\end{equation}
As the wave amplitude increases, the varying mean radius $r_*$ decreases until it reaches its minimum value when the tube is pinched closed, i.e. when $b = r_*$. According to Eq.\ (\ref{eq:correctedradius1}), this minimum radius is $r_{*\mathrm{min}} = \sqrt{2/3}\, r_0 \doteq 0.8165 \, r_0$. For the constant-volume approach, the procedure would be to determine the volume flow rate and hydraulic resistance for increasing values of the amplitude $b$, while simultaneously decreasing the mean radius $r_*$ according to Eq.\ (\ref{eq:correctedradius2}).
%(For a plane two-dimensional channel, there is no volume change caused by the sinusoidal displacement, so no such correction is necessary.) 

\subsection{The effect of the wavy wall on hydraulic resistance}
%Suppose that we compute the hydraulic resistance $\mathcal{R}$ (per unit length) of a circular tube with a wavy wall of the form (\ref{eq:sinusoid}), and we wish tocompare it with a reference uniform tube without a wavy wall. One such comparison would be to consider the reference tube to be one with constantradius $r_0$ equal to the mean radius of the wavy tube. In this case, the reference tube has hydraulic resistance 

\begin{figure}
  \centerline{\includegraphics[width=\textwidth]{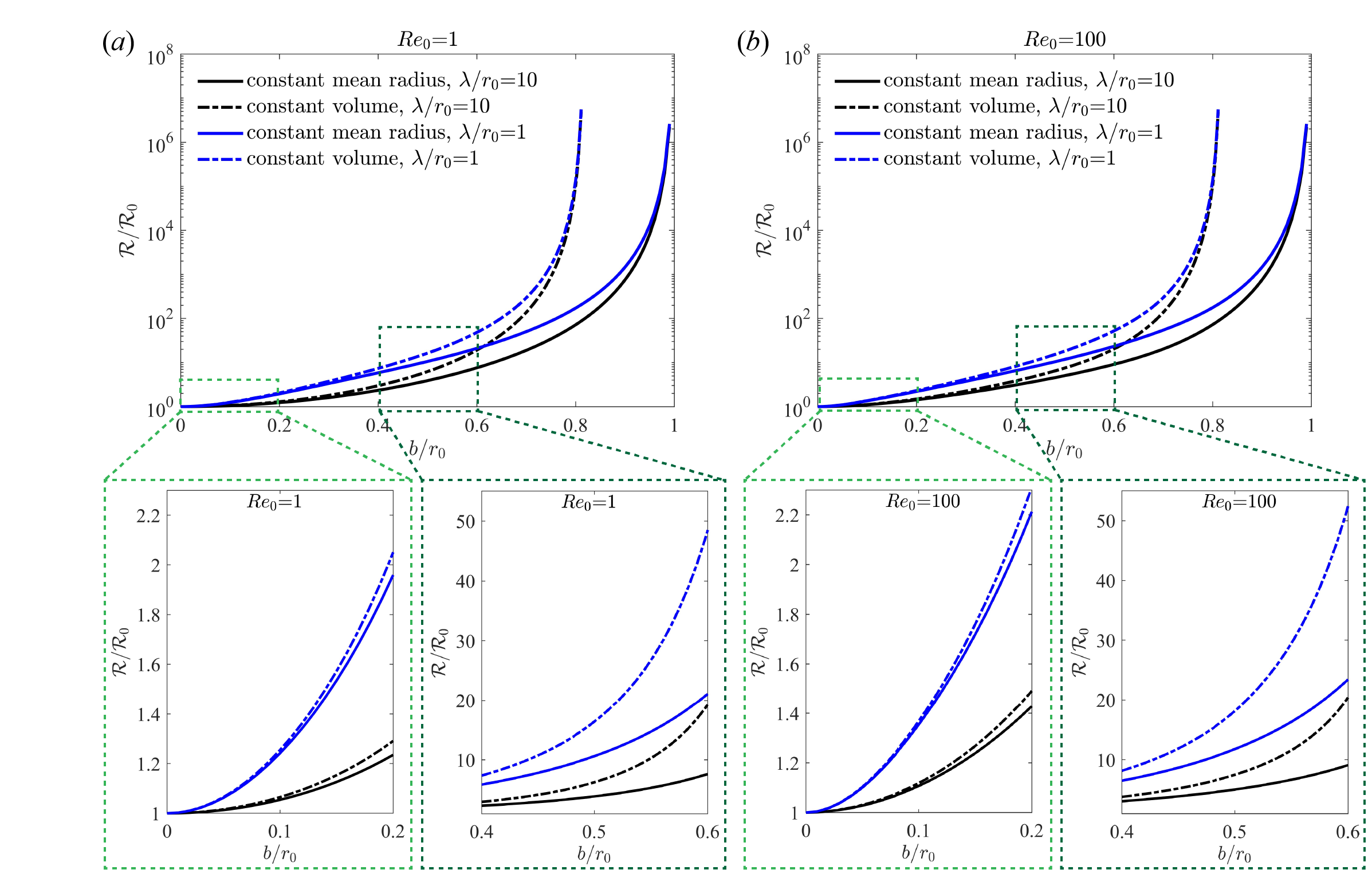}}
  \caption{Hydraulic resistance of a wavy circular tube. (a-b) Plots of scaled hydraulic resistance $\mathcal{R}/\mathcal{R}_0$ versus scaled wave amplitude $b/r_0$ for both the constant mean radius and constant volume cases, for two different reference Reynolds numbers and two different wavelengths. The driving axial pressure drop $\Delta p$ is kept constant for each plot. The hydraulic resistance is always greater in the constant-volume case, and it goes asymptotically to infinity as the scaled wave amplitude approaches its limiting value $ b/r_0 = r_{*\mathrm{min}}/r_0 = \sqrt{2/3} \doteq 0.8165$ (when the tube is pinched off into a sequence of separate compartments). } 
 \label{fig:R}
\end{figure}

%\begin{figure}
%  \centerline{\includegraphics[width=\textwidth]{./figures/Re1_surf_plot.png}}
%  \caption{Hydraulic resistance of a wavy circular tube. The Reynolds number is 1 when the tube is straight ($b/r_0 = 0$)/, and the pressure drop driving the flow is kept constant for all simulations. (a) Scaled hydraulic resistance $\mathcal{R}/\mathcal{R}_0$ is plotted against scaled wavelengths $\lambda/r_0$ and scaled wave amplitudes $b/r_0$ for both the constant mean radius and constant volume cases. The resistance is always greater in the constant-volume case, and it goes asymptotically to infinity as the scaled wave amplitude approaches its limiting value $ b/r_0 = r_{*\mathrm{min}}/r_0 = \sqrt{2/3} \doteq 0.8165$ (when the tube is pinched off into a sequence of compartments). For $\lambda/r_0 > 2$, the wavelength has little effect on $\mathcal{R}/\mathcal{R}_0$. (b) Difference in scaled resistance $\mathcal{R}/\mathcal{R}_0$ between the two cases is plotted against scaled wavelengths $\lambda/r_0$ and scaled wave amplitudes $b/r_0$.} 
% \label{fig:Re1}
%\end{figure}

\noindent To assess the effect of the volume change on hydraulic resistance, we have computed some examples of steady, pressure-driven viscous flow along a wavy tube, for the two cases: constant mean radius and constant volume. (Details of the computational method are given in the Appendix.) In particular, we are interested in the difference in the dependence of the hydraulic resistance on wave amplitude in the two cases. We define a reference Reynolds number $Re_0$ for flow in the uniform tube $(b=0)$ as $Re_0 \equiv \rho Ur_0/\mu$, where $U = Q_0/\pi r_0^2$ is the mean axial velocity over a cross-section in the undeformed tube and $\rho$ is the fluid density. Thus, 
\begin{equation} 
Re_0 \equiv \frac{\rho Q_0}{\pi \mu r_0} = \frac{\rho r_0^3}{8 \mu^2}\left(\frac{\Delta p}{L}\right),
\end{equation}
and we see that this Reynolds number can be considered as a dimensionless measure of the pressure difference $\Delta p$.
We also define a reference hydraulic resistance $\mathcal{R}_0$ to be that of the undeformed tube ($b=0$), i.e.
\begin{equation}
\mathcal{R}_0 = \frac{8 \mu}{ \pi r_0^4} . 
\end{equation}

\begin{figure}
  \centerline{\includegraphics[width=1.1\textwidth]{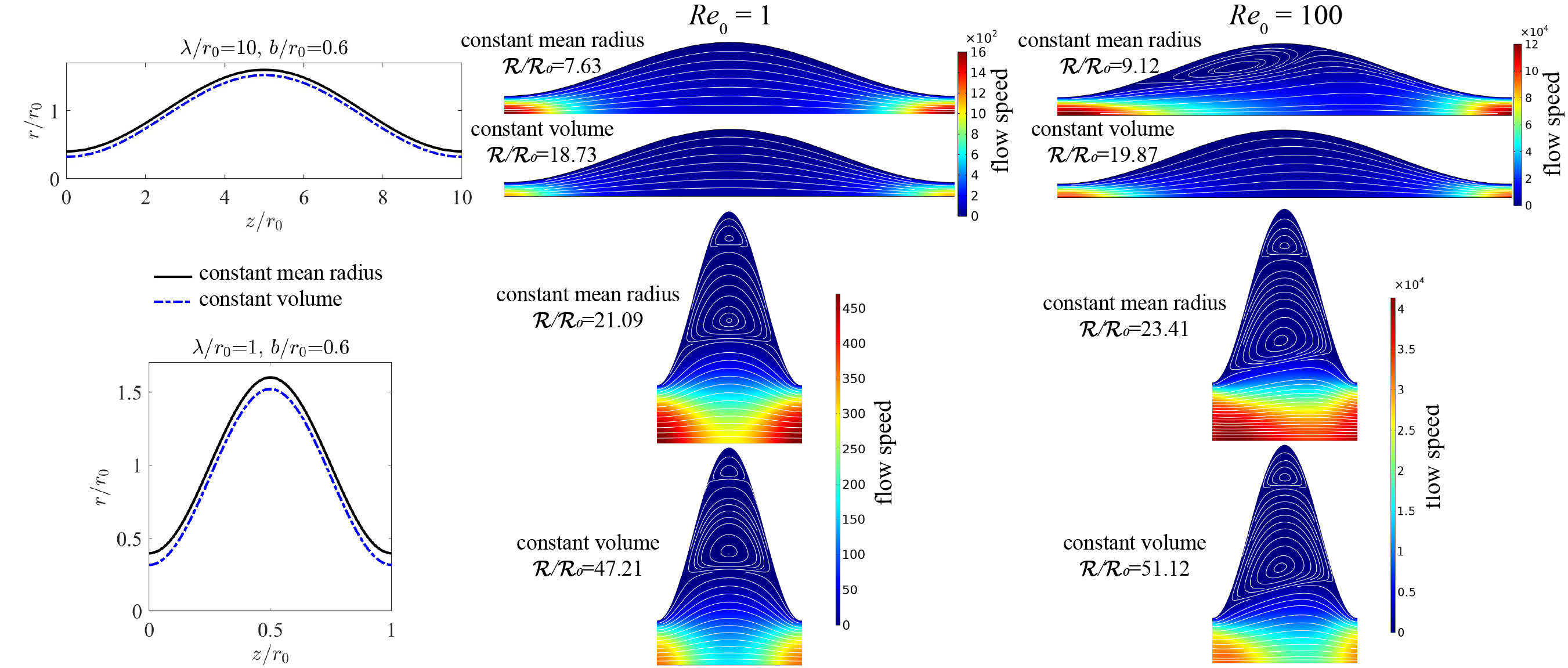}}
  \caption{Examples of the velocity field for steady Poiseuille flow in tubes with wavy walls, for the cases with constant mean radius and constant volume, for two different wavelengths ($\lambda/r_0 =$ 1 and 10), two different reference Reynolds numbers ($Re_0 =$ 1 and 100), and wave amplitude $b/r_0 = 0.6$. The left column shows the tube wall profiles, and the color plots show the streamlines, flow speed, and values of the scaled hydraulic resistance $\mathcal{R}/\mathcal{R}_0$. Note especially the case $Re_0 =$ 100, $\lambda/r_0 =$ 10, where there is a recirculating eddy in the constant-mean-radius case but not in the constant-volume case.}
  \label{fig:streamlines}
\end{figure}

In Fig.\ \ref{fig:R} we show the dependence of the normalized hydraulic resistance $\mathcal{R}/\mathcal{R}_0$ on wave amplitude for two different values of the reference Reynolds number. Here we keep the pressure difference $\Delta p$ constant while varying the wave amplitude $b$. The hydraulic resistance in the constant-volume case is systematically higher than in the constant-mean-radius case because of the decrease in mean radius required to keep the volume constant: the tube is on average narrower when the volume is kept constant. In the constant-volume case, the hydraulic resistance goes asymptotically to infinity as the scaled wave amplitude $b/r_0$ approaches its limiting value $ b/r_0 = r_{*\mathrm{min}}/r_0 = \sqrt{2/3} \doteq 0.8165$ (when the tube is pinched off into a sequence of compartments). From the enlargements in Fig.\ \ref{fig:R}(b), as the wave amplitude increases from zero, we begin to see noticeable differences in the hydraulic resistance between the constant-mean-radius and constant-volume cases for values of $b/r_0$ as small as 0.1. The differences increase asymptotically as $b/r_0$ approaches 1 and become quite significant at larger wave amplitudes (for example, see the values in Figure \ref{fig:streamlines} for the cases with $b/r_0=0.6$). For each reference Reynolds number, we show results for a small wavelength ($\lambda/r_0=1$) and a large wavelength ($\lambda/r_0$=10). The differences between these two cases show that shorter wavelengths produce significantly greater hydraulic resistance. 

%As an alternative scaling for the constant volume case, we could consider the reference uniform tube to be one with the same volume as the wavy tube. In this case, the reference tube has radius $r_*$, which varies with wave amplitude $b$, and varying hydraulic resistance
%\begin{equation}
%\mathcal{R}_{*} = \frac{8 \mu}{\pi r_*^4} = \frac{8 \mu}{\pi r_0^4} \left[ 1 + \frac{1}{2}\left(\frac{b}{r_0}\right)^2 %\right]^2 .
%\end{equation}
%This reference value of the hydraulic resistance depends on the wave amplitude $b$ and is greater than $\mathcal{R}_0$ because the mean radius $r_*$ is smaller than $r_0$ in order to keep the volume unchanged. %An example of the dependence of the scaled hydraulic resistance $\mathcal{R}/\mathcal{R_*}$ on the scaled wave amplitude $b/r_*$ is shown in Figure 2. 

In order to illustrate the differences in the flows represented in Fig.\ \ref{fig:R}, in 
Fig.\ \ref{fig:streamlines} we show the detailed flow speeds, streamline patterns, and scaled hydraulic resistances for a particular value of the scaled wave amplitude, $b/r_0 = 0.6$. Here we see that the hydraulic resistances of the corresponding constant-volume and constant-mean-radius cases differ by more than a factor of two. An important feature of these flows is the formation of recirculating eddies in the wide parts of the tube at a critical value of the wave amplitude. Once the recirculating eddies have formed, the axial flow is effectively confined to a channel much narrower than the mean radius, thus producing greatly increased hydraulic resistance. Note especially, in the case $Re_0 = 100, \lambda/r_0 = 10$, that a recirculating eddy has formed in the constant-mean-radius case but not in the constant-volume case: here the two flows are not just quantitatively different, they are also qualitatively different.

\subsection{Dimensional analysis of the flow}
We can relate the results for the constant-volume case to those of the constant-mean-radius case through a suitable scale analysis.
Consider a finite section of wavy tube with circular cross section, of length $L$ (an integer multiple of $\lambda$), with pressures $p_0$ and $p_L$ at the entrance and exit. For fully-developed viscous flow along this tube, the relevant dimensional parameters are the volume flow rate $Q$, the pressure drop $\Delta p = p_0 - p_L$, the viscosity $\mu$, the density $\rho$, the radius $r_0$ of the undeformed tube, the length $L$ of the tube, and the amplitude $b$ and wavelength $\lambda$ of the sinusoidal displacement. According to the Buckingham $\pi$ theorem, these eight dimensional quantities can be arranged into five dimensionless quantities: 
\begin{equation}
\hat{Q} \equiv Q/Q_0, \quad 
     Re_0 \equiv \frac{\rho Q_0}{\pi \mu r_0} = \frac{\rho r_0^3}{8 \mu^2}\left(\frac{\Delta p}{L}\right), \quad \frac{L}{r_0}, \quad \frac{b}{r_0}, 
     \quad \frac{\lambda}{r_0} .
\label{eq:scaling1}
\end{equation}
Here $Q_0 = \pi r_0^4 (\Delta p/L)/8\mu$ is the volume flow rate in the undeformed tube (for $b=0$) and $\hat{Q}$ is a dimensionless measure of the volume flow rate in the wavy tube, scaled so that $\hat{Q} = 1$ for the undeformed tube. $Re_0$ is the Reynolds number based on the mean axial flow velocity $Q_0/\pi r_0^2$ in the undeformed tube; note that $Re_0$ can also be considered as a dimensionless measure of the pressure difference $\Delta p$. The flow is then described by a functional relationship of the form 
\begin{equation}
    \hat{Q} = \mathcal{F}(Re_0, \frac{L}{r_0}, \frac{b}{r_0}, \frac{\lambda}{r_0})
\label{eq:scaling2}
\end{equation}
(Note that the mean axial pressure gradient is given by $\overline{\partial p / \partial z} = - \Delta p/ L$.)

%For the case we consider in this paper, fully developed flow in long tubes, the flow is independent of $L/r_0$. In this case, the dimensionless flow rate $\hat{Q}$ depends only on the Reynolds number $Re$ and the geometric ratios $b/r_0$ and $\lambda/r_0$.

For the case in which the volume is kept constant, we can define the following dimensionless quantities, using the varying mean radius $r_*$ as the reference radius:
\begin{equation}
\hat{Q}_* \equiv \frac{Q_*}{Q_{*0}}, \quad 
     Re_{*0} \equiv \frac{\rho Q_{*0}}{\pi \mu r_*} = \frac{\rho r_*^3}{8 \mu^2}\left(\frac{\Delta p_*}{L_*}\right), \quad \frac{L_*}{r_*}, \quad \frac{b_*}{r_*}, 
     \quad \frac{\lambda_*}{r_*},
\label{eq:scaling3}
\end{equation}
where $Q_{*0} = \pi r_*^4 (\Delta p_*/L_*)/8\mu $ is the volume flow rate in a uniform tube of radius $r_*$.
The flow is then described by a functional relationship of the form 
\begin{equation}
    \hat{Q_*} = \mathcal{F}(Re_{*0}, \frac{L_*}{r_*}, \frac{b_*}{r_*}, \frac{\lambda_*}{r_*})
\label{eq:scaling4}
\end{equation}
For the undeformed tube ($b=0$), with $r_* = r_0$, the two scalings are the same. (In place of the volume flow rate, we could consider the hydraulic resistance $\mathcal{R_*} = (\Delta p_*/L_*)/Q_*$ as the dependent variable, scaled by $\mathcal{R}_0$.) 

\begin{figure}
  \centerline{\includegraphics[width=\textwidth]{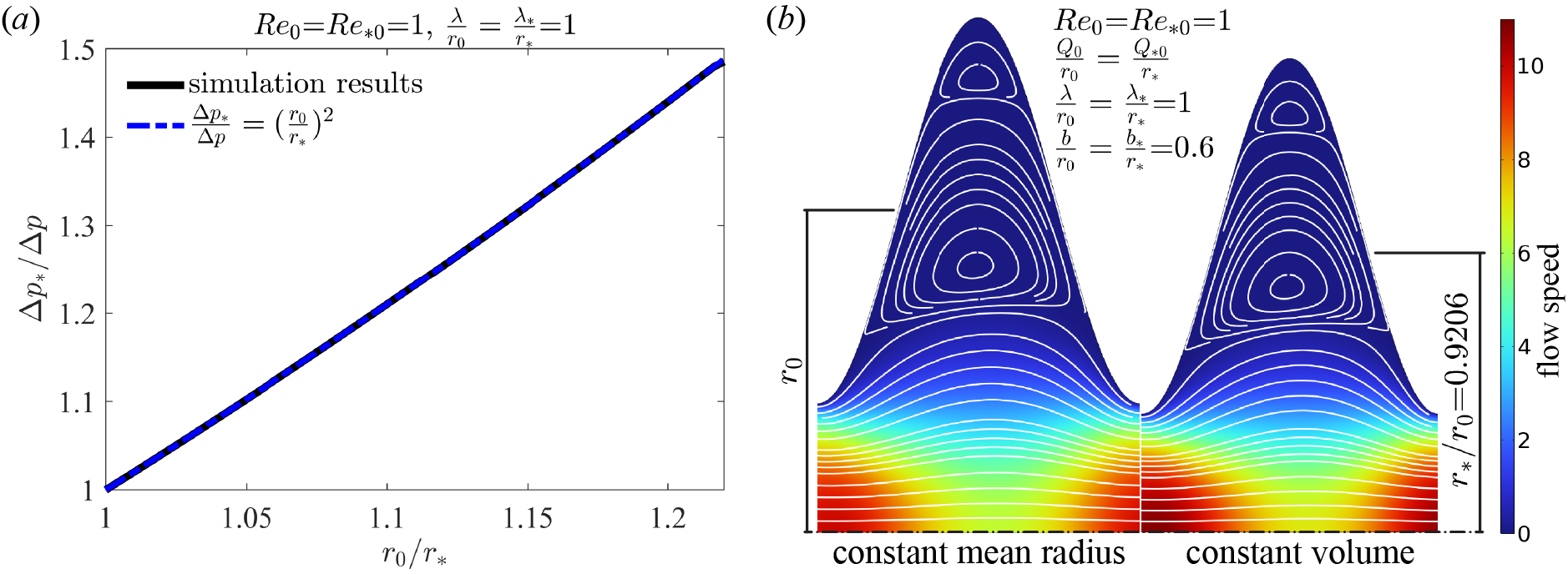}}
  \caption{Numerical simulations with equal scaled volume flow rates $\hat{Q}$ and $\hat{Q_*}$ ($Re_{0}$ = $Re_{*0}$ = 1, $\frac{Q_0}{r_0} = \frac{Q_{*0}}{r_*}$, $\frac{\lambda}{r_0} = \frac{\lambda_*}{r_*}$=1, and $\frac{b}{r_0} = \frac{b_*}{r_*}$). (a) To maintain dynamic similarity with the constant-mean-radius case, the pressure drop $\Delta p_*$ in the constant-volume case increases by a factor of $(r_0/r_*)^2$ as the mean radius $r_*$ decreases. (b) For $\frac{b}{r_0} = \frac{b_*}{r_*}$ = 0.6, the constant-mean-radius and constant-volume cases show similar flow fields.}
  \label{fig:dimensional_analysis}
\end{figure}

The functional relationships $\mathcal{F}$ expressed in Eqs.\ (\ref{eq:scaling2}) and (\ref{eq:scaling4}) are the same and can be used to determine properties of constant-volume flows from those of constant-mean-radius flows. In order to have dynamic similarity between the two cases, assuming the same fluid properties, we require geometric similarity, $L/r_0 = L_*/r_*$, $b/r_0 = b_*/r_*$, $\lambda/r_0 = \lambda_*/r_*$, and equal Reynolds numbers, $Re_0 = Re_{*0}$, that is, require that $Q_0/r_0 = Q_{*0}/r_*$, or $r_0^3 \Delta p/L$ = $r_*^3 \Delta p_*/L_*$. Hence, we have general dynamic similarity between the two cases if we scale the pressure drop in the constant-volume case as 
\begin{equation}
    \Delta p_* = (\frac{r_0}{r_*})^3 \frac{L_*}{L} \Delta p ,
\end{equation}
or, since $L_*/L = r_*/r_0$,
\begin{equation}
    \Delta p_* = (\frac{r_0}{r_*})^2 \Delta p .
\end{equation}
Thus, for dynamic similarity, the pressure drop $\Delta p_*$ in the constant-volume case must be increased by a factor of $(r_0/r_*)^2$ as the mean radius $r_*$ decreases. We confirmed this relationship through simulations that kept the scaled volume flow rates $\hat{Q}$ and $\hat{Q_*}$ equal. For $Re_0 = 1$ and $\lambda/r_0 = 1$, the simulations show that the ratio $\Delta p_*/\Delta p$ is indeed proportional to $(r_0/r_*)^2$, from $r_0/r_* = 1$ up to $r_0/r_* \approx 1.217$, just before the tube pinched off, as shown in Fig.\ \ref{fig:dimensional_analysis}(a). Figure \ref{fig:dimensional_analysis}(b) shows, for example, that when $\frac{b}{r_0} = \frac{b_*}{r_*}$=0.6, by keeping the scaled volume flow rates $\hat{Q}$ and $\hat{Q_*}$ equal, the constant-mean-radius and constant-volume simulation cases have similar flow fields, even though $r_*$ is much smaller than $r_0$.

The functional relationship in Eq.\ (\ref{eq:scaling2}) for the constant-mean-radius case has been determined numerically, for various ranges of values of the dimensionless parameters, in several published studies (including those cited here). Based on these studies, or further numerical simulations, the equivalent functional relationship (\ref{eq:scaling4}) can be used to determine the flow properties in the constant-volume case in corresponding ranges of parameters. 
%For example, Fig.\ \ref{fig:dimensional_analysis}(b) shows that when $\frac{b}{r_0} = \frac{b_*}{r_*}$=0.6, by keeping the scaled volume flow rates $\hat{Q}$ and $\hat{Q_*}$ equal, the constant-mean-radius and constant-volume simulation cases produce similar flow fields, even though the $r_*$ is much smaller than $r_0$.

\subsection{Peristaltic pumping}

\noindent Peristaltic pumping is the driving of a net (time-averaged) flow along a tube caused by a wall wave propagating in the axial direction. Theoretical studies of peristaltic pumping in a circular tube usually assume a sinusoidal wall wave of the same form as Eq. (\ref{eq:sinusoid}), but propagating in the $z$-direction, i.e.:
\begin{equation}
r(z,t) = r_0 + b \sin\left[ \frac{2\pi}{\lambda}(z - ct) \right] ,
\label{eq:propwave}
\end{equation}
where $c$ is the wave speed.
These studies often consider the flow driven by both the wall motion and a constant pressure drop along the length of the tube, in which case the flow due to peristaltic pumping alone can be found by setting the pressure drop equal to zero. 
For the propagating sinusoidal wall displacement (\ref{eq:propwave}), the net volume change is the same as that discussed above. Thus, in evaluating the increase in the mean flow rate due to increasing amplitude $b$ of the wall wave, three effects come into play: the greater pushing effect of the wave on the fluid, the change in hydraulic resistance, and the increase in the volume of fluid inside the tube. Previous studies focus only on the first two of these effects. Here, we briefly consider the effects of the volume change, which are considerable. 

For peristaltic pumping in a circular tube with a propagating sinusoidal wave of the form (\ref{eq:propwave}), the time-averaged volume flow rate $\overline{Q}$ is given by \cite{shapiro1969, jaffrin1971}
\begin{equation}
    \overline{Q} = 8 \pi r_0^2 c \,\left(\frac{b}{r_0}\right)^2 \left(\frac{1 - \frac{1}{16}(\frac{b}{r_0})^2}{2 + 3 (\frac{b}{r_0})^2}\right) .
\label{eq:peristalsis1}
\end{equation}
Here we see that, for small wave amplitudes $b/r_0 << 1$, the mean flow rate due to peristaltic pumping scales very nearly as $(b/r_0)^2$, which is the same scaling as the relative volume change, so the volume change will always be a significant effect in the context of peristalsis, no matter how small the wave amplitude. 
Because peristaltic pumping often involves large relative amplitudes of the wall wave, the volume change is of considerable importance. For example,
in the extreme case where $b = r_0$ and the interior of the tube is pinched off into separate compartments, all of the fluid in the tube will be forced to move downstream at the wave speed $c$,
and the volume flow rate $\overline{Q}$ will have its maximum value (cf.\ Eq.\ \ref{eq:peristalsis1}),
\begin{equation}
\overline{Q}_{\rm max} = \frac{3}{2} \pi r_0^2 c = \frac{V_{\lambda {\rm max}}}{\lambda/c}.
\label{eq:peristalsis2}
\end{equation}
If instead we keep the volume constant by reducing the mean radius and rescaling the amplitude of the wall wave, as described above, we obtain the limiting value $\overline{Q}_{\rm max} = \pi r_0^2 c$,
which corresponds to the case where the entire volume of fluid contained in the tube (the same volume as in the undeformed tube) is moving downstream at speed c.
Hence, we see that the maximum volume flow rate is 50\% greater if we keep the mean radius, rather than the volume, constant while increasing the wave amplitude, a difference that is solely due to the difference in volume in these two cases.

\section{Annular tubes}

\subsection{The annular tube with a sinusoidal outer boundary}
\noindent Here we consider a concentric circular annular tube of inner radius $r_1$ and mean outer radius $r_2$, with a sinusoidal displacement superposed on the outer wall, i.e.,
	\begin{equation}
	r(z) = r_2 + b\sin(2\pi z/\lambda) ,
	\end{equation}
and the cross-sectional area $A(z)$ is
	\begin{equation}
	A(z) = \pi [(r_2 + b\sin(2\pi z/\lambda))^2 -r_1^2] = \pi [r_2^2 + 2 r_2 b \sin(2\pi z/\lambda) 
	+ b^2 \sin^2(2\pi z/\lambda) - r_1^2] .
	\end{equation}
The interior volume $V_\lambda$ over one wavelength along the tube is then given by 
	\begin{equation} 
	V_\lambda = \int_0^\lambda A(z) dz = \pi [ (r_2^2 - r_1^2) + \frac{b^2}{2} ] \lambda ,
	\end{equation} 
and hence the volume per wavelength is increased by the same amount, $\pi b^2 \lambda/2$, as for the circular tube (cf.\ Eq.\ (\ref{eq:volume1})). This is an obvious result because the inner wall is not displaced and hence contributes nothing to the volume change: letting $r_1 \rightarrow 0$ and setting $r_2 = r_0$, we obtain the results in the previous section for the open tube. Note, however, that here the maximum possible value of the wave amplitude is $b = r_2 - r_1$, so the maximum increase in volume $V_\lambda$ is $(r_2 - r_1)^2 /2$. 

\begin{figure}
  \centerline{\includegraphics[width=\textwidth]{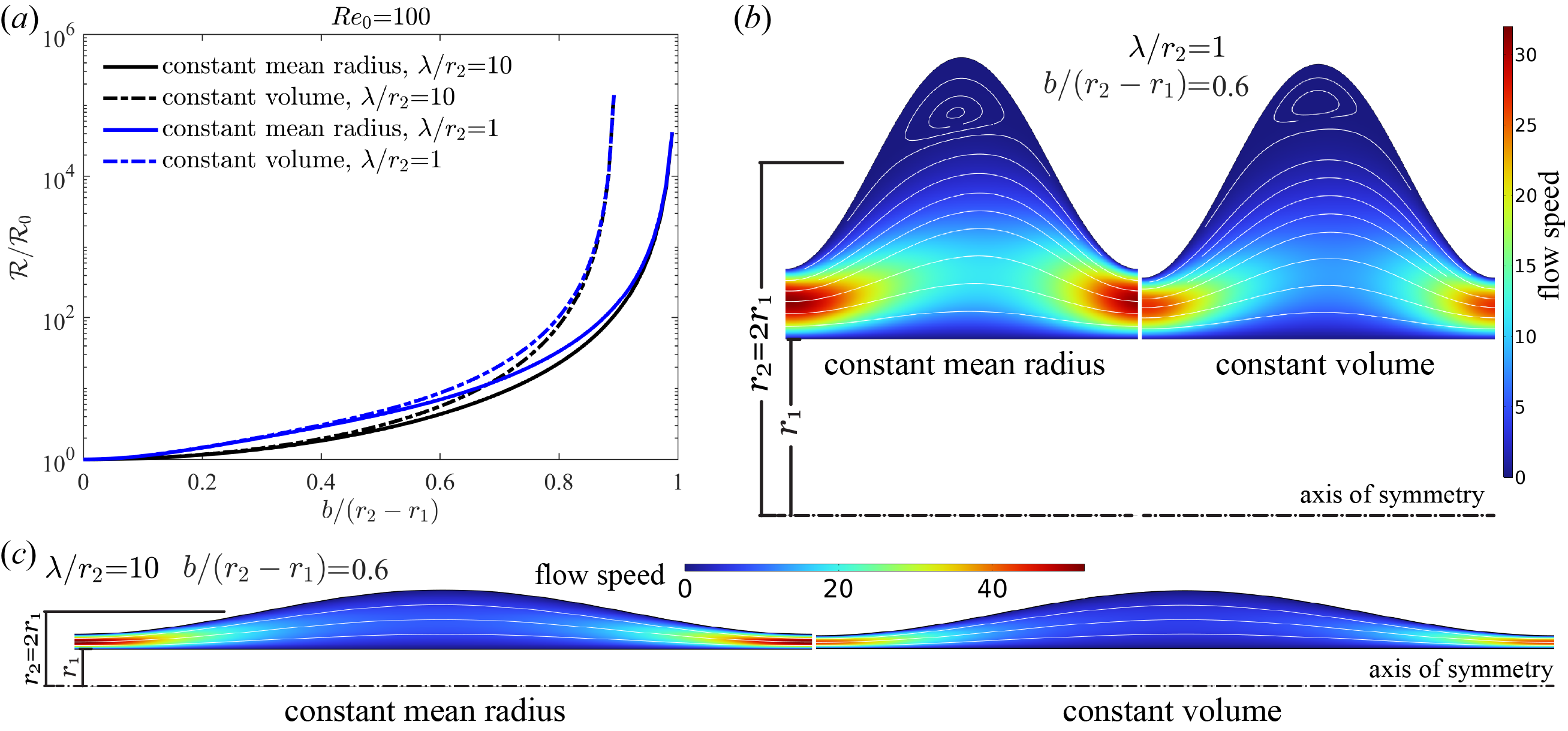}}
  \caption{Annular tube with sinusoidal outer boundary. (a) Scaled hydraulic resistance $\mathcal{R}/\mathcal{R}_0$ versus scaled wave amplitude $b/(r_2-r_1)$ for $Re_0 = 100$, $r_2 = 2r_1$, and $\lambda/r_2 = 1$ and $10$. (b-c) Plots of flow speeds for the constant-mean-radius and constant-volume cases, for $b/(r_2-r_1) = 0.6$.} 
 \label{fig:annulus_outer}
\end{figure}

Consider, for example, the case where $r_2 = 2 r_1$, for which the volume increases from $V_\lambda = \frac{3}{4} \pi r_2^2$ (for $b=0$) to $V_\lambda = \frac{7}{8} \pi r_2^2$ (for $b = b_{\rm max} = r_2-r_1 = r_2/2$). Figure \ref{fig:annulus_outer}(a) shows the scaled hydraulic resistance, $\mathcal{R}/\mathcal{R}_0$, as a function of the scaled wave amplitude $b/(r_2-r_1)$ in the range from 0 to maximum for this example with $Re_0 = 100$ and $\lambda/(r_2-r_1) = 1$ and $10$. Figures \ref{fig:annulus_outer}(b-c) compare the flow speeds for $b/(r_2-r_1) = 0.6$, and the constant-volume case shows decreased flow speed. Consistent with the results in the open tube, the hydraulic resistance in the constant-volume case is higher than in the constant-mean-radius case, and shorter wavelengths produce significantly greater hydraulic resistance.

\subsection{The annular tube with a sinusoidal inner boundary}
\noindent If we again consider a concentric circular annular tube, but with the sinusoidal displacement superposed on the inner wall, we have the following:
	\begin{equation}
	A(z) = \pi [r_2 ^2 - (r_1+ b\sin(2\pi z/\lambda))^2] = \pi [r_2^2 - r_1^2 - 2 r_1 b \sin(2\pi z/\lambda) 
	- b^2 \sin^2(2\pi z/\lambda) ],
	\end{equation}
	\begin{equation}
	V_\lambda = \int_0^\lambda A(z) dz = \pi [ (r_2^2 - r_1^2) - \frac{b^2}{2} ] \lambda .
	\end{equation} 
Thus, in this case the volume per wavelength in the tube is reduced by the amount $\pi b^2 \lambda /2$. In order to eliminate the effect of the decreasing volume on volume flow rate for pressure-driven or peristaltic-driven flow, one could keep the volume constant by decreasing the mean radius of the inner wall as the amplitude of the wall wave increases, in a manner analogous to that discussed above for the open tube. 

\begin{figure}
  \centerline{\includegraphics[width=\textwidth]{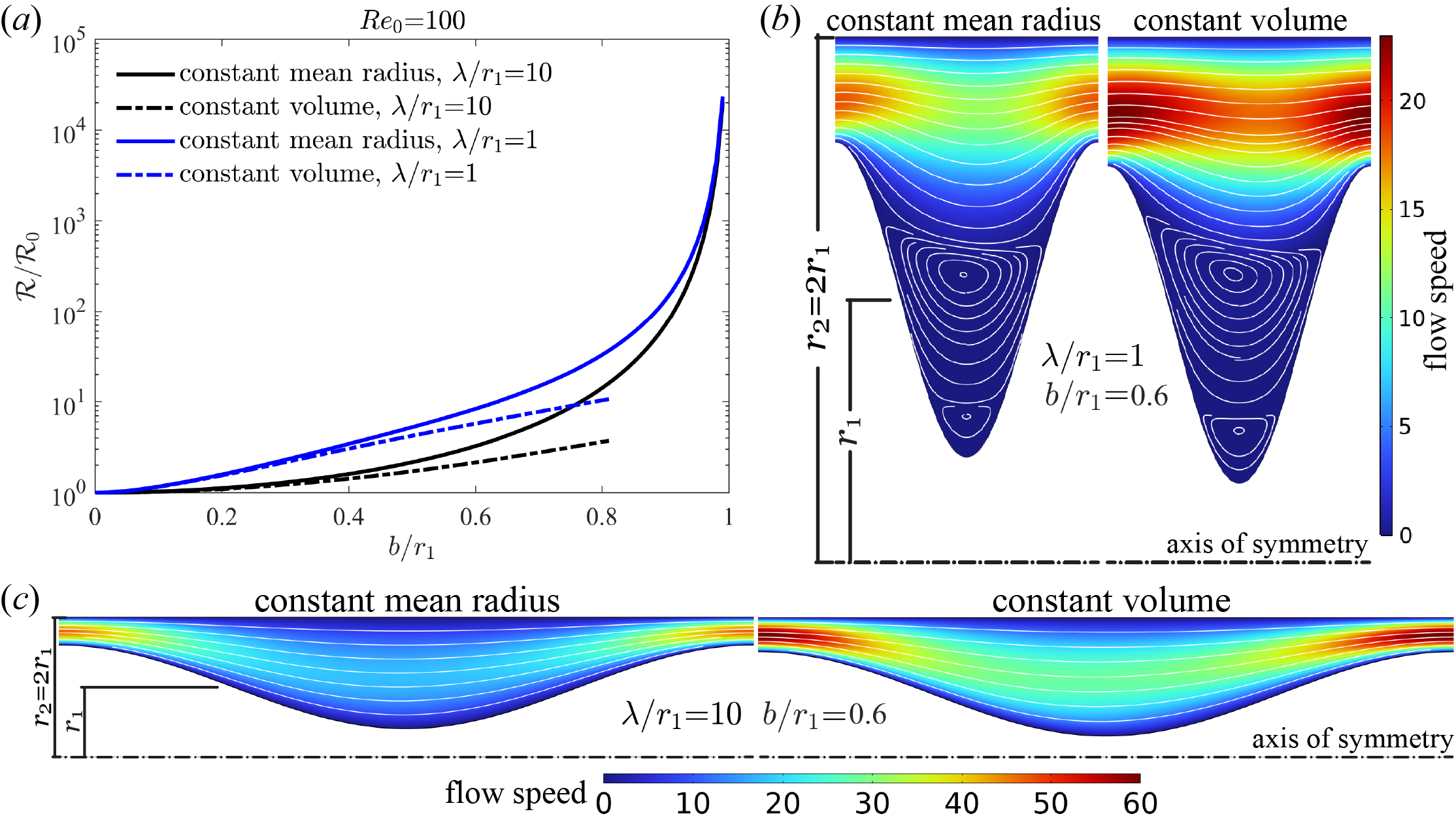}}
  \caption{Annular tube with sinusoidal inner boundary. (a) Scaled hydraulic resistance $\mathcal{R}/\mathcal{R}_0$ versus scaled wave amplitude $b/r_1$ for $Re_0 = 100$, $r_2 = 2r_1$, and $\lambda/r_1 = 1$ and $10$. (b-c) Plots of flow speeds for the constant-mean-radius and constant-volume cases, for $b/r_1 = 0.6$.} 
 \label{fig:annulus_inner}
\end{figure}

We show the example with $r_2 = 2r_1$ in Fig.\ \ref{fig:annulus_inner}. The scaled hydraulic resistance, $\mathcal{R}/\mathcal{R}_0$, is lower in the constant-volume case, especially at higher values of the scaled wave amplitude $b/r_1$. For flows in the open tube and the annular tube with a sinusoidal outer boundary, the constant-mean-radius case underestimates the hydraulic resistance. Conversely, for flow in the annular tube with a sinusoidal inner boundary, constant-mean-radius case overestimates the hydraulic resistance because of the volume decrease in the tube. At $Re_0 = 100$ and $\lambda/r_1 = 1$, the maximum overestimation reaches approximately 340\%.

It is worth noting that for an annular tube with either a wavy outer or wavy inner wall, or both together, the volume change is the same if the annulus is eccentric, that is, if the axis of the inner tube is not coincident with the axis of the outer tube. (Of course the maximum possible amplitude of the wall wave, when the inner and outer walls are just touching, becomes smaller as the eccentricity increases.) Although the volume change remains the same, the hydraulic resistance will decrease with increasing eccentricity because the decrease in shear stress in the expanded parts of the annulus outweighs the increase in shear stress in the narrowed parts \cite{white2006, tithof2019}.

For peristaltic pumping in an annular tube with a propagating sinusoidal wave (of the form (\ref{eq:propwave})) of small amplitude $b$ on the inner wall, the time-averaged volume flow rate $\overline{Q}$ is given by \cite{wang2011}
\begin{equation}
    \overline{Q} = \pi r_1^2 c \,\left(\frac{b}{r_1}\right)^2 \left(\frac{2(\frac{r_1}{r_2})^2}{1-(\frac{r_1}{r_2})^2}\right) .
\label{peristalsis3}
\end{equation}
We note that $\overline{Q}$ scales as $(b/r_1)^2$ for small wave amplitude $b$, the same scaling as for the relative volume change, so, as in the case of the circular tube discussed above, the volume change due to the wavy wall is always a significant effect, even for a very small wave amplitude. The formula (\ref{peristalsis3}) was obtained in a study of peristaltic pumping in perivascular spaces in the brain due to arterial pulsations \cite{wang2011}.

\section{Discussion}

\noindent We have shown that the usual way of introducing a wavy displacement of a wall of a circular or annular tube, keeping the mean radius constant, produces a change in volume that has significant effects that have been generally ignored. For laminar viscous flow along a wavy circular tube, we find that keeping the interior volume, rather than the mean radius, constant leads to greater hydraulic resistance, and in some cases produces flows that are qualitatively different, in the sense of the presence or absence of recirculating eddies. We have provided a dimensional scaling analysis that relates the volume flow rate in the constant-volume case to that in an equivalent constant-mean-radius case. The volume change is especially significant for peristaltic pumping by a moving wall wave, producing up to 50\% greater volume flow rate than in the case where the volume is held constant.

Here we have considered only a sinusoidal displacement of the tube wall. Alternatively, one could specify a different, non-sinusoidal form of periodic displacement that preserves the interior volume without changing the mean radius: for an open tube, such a displacement would have a smaller amplitude in the outward direction than in the inward direction. However, in specifying such a displacement, we would lose the mathematical simplicity and generality of the purely sinusoidal form. 

Our interest in this problem arose from our research on the flow of cerebrospinal fluid in perivascular spaces (PVSs) surrounding arteries and arterioles in the brain \cite{bohr2022, kelley2023}. This flow is largely driven by arterial pulsations associated with the heartbeat, for which the waveform is not purely sinusoidal, but experimental data on blood vessel wall motion \cite{mestre2018} demonstrates volume conservation, consistent with expectation that blood volume is closely regulated in the closed circulatory system. The relative amplitude of these pulsations is of order 1\%, but slower pulsations associated with functional hyperemia have relative amplitudes of order 10\%. 

The volume-change effects studied here are relevant to the problem of peristaltic pumping mechanisms in annular PVS geometries \cite{wang2011, carr2021, coenen2021} and also to the effect of constrictions along a PVS on its hydraulic resistance \cite{raicevic2023, boster2024}. For example, when modeling fluid flow in the PVS surrounding a blood vessel with a sinusoidal wall displacement, assuming the PVS thickness is 15\% of the vessel radius with parameters $b/r_0 = 0.1$ and $\lambda/r_0 = 10$, the constant-mean-radius method underestimates the hydraulic resistance by approximately 20\%.

\begin{acknowledgments}
We thank our colleague Doug Kelley for helpful suggestions. This research was supported by the US National Center for Complementary and Integrative Health (grant no. R01AT012312), and by the BRAIN Initiative of the US National Institutes of Health (grant no. U19NS128613).

\end{acknowledgments}

\appendix*
\section{Computational method}
\noindent We calculate steady, laminar, viscous, incompressible flows along a wavy tube by solving the governing Navier-Stokes equations
\begin{subequations}
\begin{align}
\rho (\bfu \cdot \bfnabla)\bfu &= -\bfnabla p+\mu \nabla^2 \bfu,
\\
\bfnabla \cdot \bfu &= 0,
\end{align}
\end{subequations}
using a finite element method. Here $\bfu$ is the velocity field, $p$ is the pressure, $\rho$ is the density, and $\mu$ is the dynamic viscosity.

\begin{figure}
  \centerline{\includegraphics[width=0.7\textwidth]{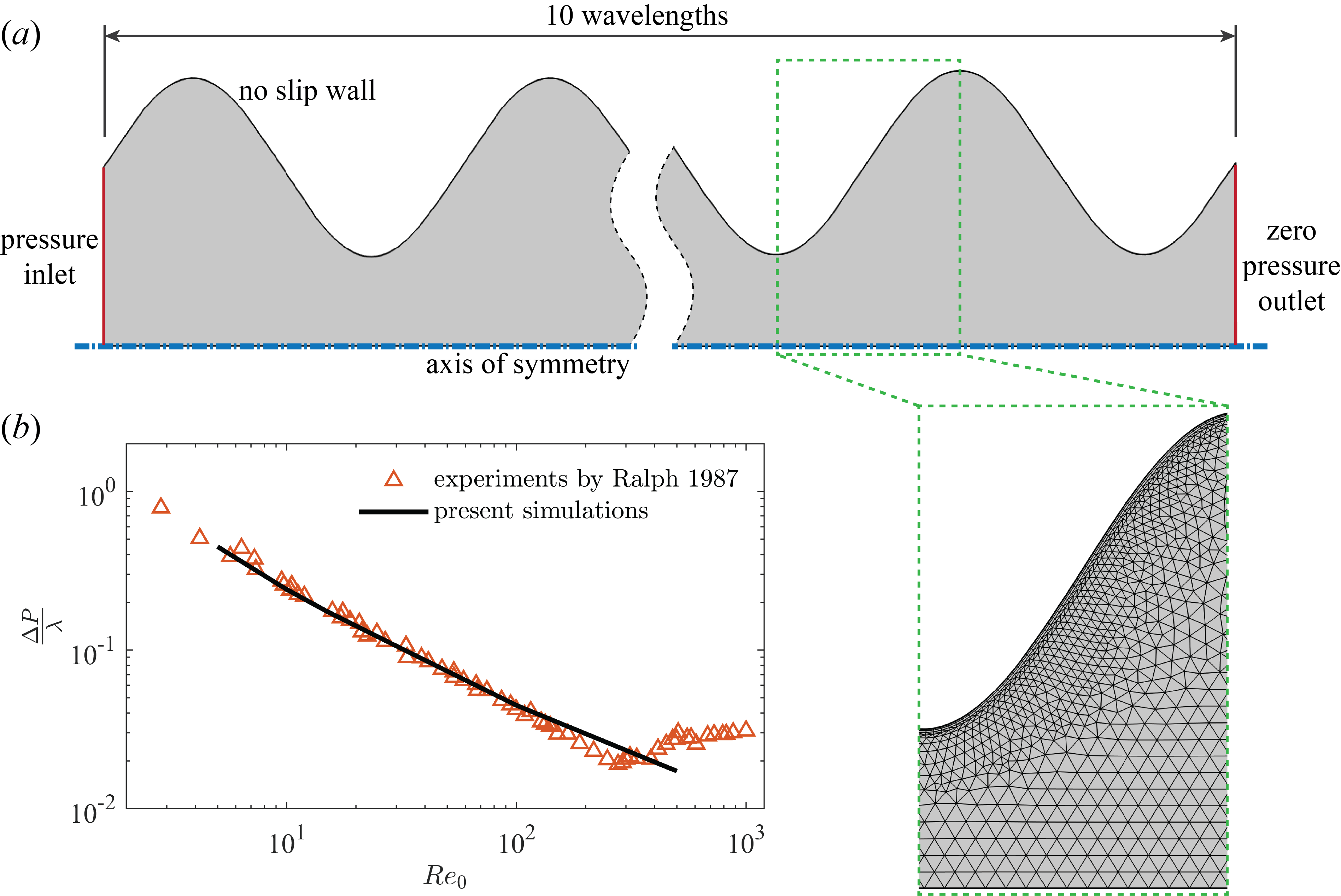}}
  \caption{(a) The computational domain and the finite-element mesh. (b) Comparison of present simulation results with experimental measurements of pressure drop per wavelength vs. Reynolds number $Re_0$ by \cite{ralph1987}.}
  \label{fig:appendix}
\end{figure}

A sketch of the configuration of the computational domain is given in Figure \ref{fig:appendix}(a). Simulations of two-dimensional, axisymmetric, steady flow were carried out using the finite-element code COMSOL Multiphysics 6.3. At the inlet, the flow is fully developed flow in a uniform tube, which is calculated by assuming that the fluid enters the computational domain through a virtual straight tube, an extrusion of the inlet cross section of length ten times radius $r_0$. An example of the finite-element mesh is shown in Figure \ref{fig:appendix}(a). A mesh sensitivity study was first performed to ensure that the meshes were sufficiently fine to resolve the computational domain. The maximum element size is $0.05r_0$, with a finer mesh near the wall, and the numerical results did not change substantially when the mesh size was further decreased.

The simulation results were validated by comparing with the experimental results of \cite{ralph1987}, who measured the pressure drop per wavelength of a sinusoidal tube with varying Reynolds number ($Re_0$), defined based on the mean tube radius and mean velocity. The tube configuration is set to be the same as in the experiments, with $\lambda = 5r_0$ and $b =0.5r_0$. As illustrated in \ref{fig:appendix}(b), our simulation results show good agreement with the experimental measurements up to the point where the flow transitions to turbulence at approximately $Re_0 = 300$.

\bibliography{JHT_Glymphatics}% Produces the bibliography via BibTeX.

@Article{bohr2022,
    author = {Bohr, T. and Hjorth, P. G. and Holst, S. C. and Hrab{\v e}tov{\'a}, S. and Kiviniemi, V. and Lilius, T. and Lundgaard, I. and Mardal. K.-A. and Martens, E. A. and Mori, Y. and N{\"a}gerl, U. V. and Nicholson, C. and Tannenbaum, A. and Thomas, J. H. and Tithof, J. and Benveniste, H. and Iliff, J. J. and Kelley, D. H. and Nedergaard, M.},
	journal = {iScience},
	number = {9},
	pages = {104987},
	title = {The glymphatic system: Current understanding and modeling},
	volume = {25},
	year = {2022}
}

@Article{boster2024,
author = {Boster, K. A. S. and Sun, J. and Shang, J. K. and Kelley, D. H. and Thomas, J. H.},
title = {Hydraulic resistance of three-dimensional pial peerivascular spaces in the brain},
journal = {Fluids Barriers CNS},
volume = {21},
number = {7},
pages = {},
year = {2024},
doi = {10.1186/s12987-023-00505-5},
}

@Article{carr2021,
  author  = {Carr, J. B. and Thomas, J. H. and Liu, J. and Shang, J. K.},
  title   = {Peristaltic pumping in thin non-axisymmetric annular tubes},
  journal = {J. Fluid Mech.},
  year    = {2021},
  volume  = {917},
  number  = {A10},
  pages   = {},
}

@Article{hemmat1995,
  author    = {Hemmat, M. and Borhan, A.},
  title     = {Creeping flow through sinusoidally constricted pipes},
  journal   = {Phys. Fluids},
  year      = {1995},
  volume    = {9},
  pages     = {2111--2121},
}

@Article{kelley2023,
author = {Kelley, D. H. and Thomas, J. H.},
title = {Cerebrospinal Fluid Flow},
journal = {Annu. Rev. Fluid Mech.},
volume = {55},
number = {1},
pages = {},
year = {2023},
doi = {10.1146/annurev-fluid-120720-011638},
URL = {},
eprint = {},
}

@Article{mestre2018,
  author  = {Mestre, H. and Tithof, J. and Du, T. and Song, W. and Peng, W. and Sweeney, A. M. and Olveda, G. and Thomas, J. H. and Nedergaard, M. and Kelley, D. H.},
  title   = {Flow of cerebrospinal fluid is driven by arterial pulsations and is reduced in hypertension},
  journal = {Nat. Commun.},
  year    = {2018},
  volume  = {9},
  number  = {1},
  pages   = {4878},
}

@Article{raicevic2023,
  author    = {Raicevic, N. and Forer, J. M. and Ladr\'on-de-Guevera, A. and Du, T. and Nedergaard, M. and Kelley, D. H. and Boster, K.},
  title     = {Sizes and shapes of periarterial spaces surrounding murine pial arteries},
  journal   = {Fluids Barriers CNS},
  year      = {2023},
  number    = {},
  pages     = {},
  volume    = {20:56},
  publisher = {},
}

@Article{tithof2019,
  author  = {Tithof, J. and Kelley, D.~H. and Mestre, H. and Nedergaard, M. and Thomas, J.~H.},
  title   = {Hydraulic resistance of periarterial spaces in the brain},
  journal = {Fluids Barriers CNS},
  year    = {2019},
  volume  = {16},
  number  = {19},
}

@Article{wang2011,
  author  = {Wang, P. and Olbricht, W. L.},
  title   = {Fluid mechanics in the perivascular space},
  journal = {J. Theor. Biol.},
  year    = {2011},
  volume  = {274},
  number  = {1},
  pages   = {52--57},
  month   = apr,
}

@Book{white2006,
  author    = {White, F. M.},
  title     = {Viscous Fluid Flow},
  publisher = {McGraw-Hill},
  address = {New York},
  year  = {2006},
  edition = {3rd}
}

@Article{belinfante1962,
	Author = {Belinfante, D. C.},
	Journal = {Proc. Cambridge Phil. Soc.},
	Pages = {405--416},
	Title = {On viscous flow in a pipe with constrictions},
	Volume = {58},
	Year = {1962}
}

@Article{burns1967,
  author  = {Burns, J. C. and Parks, T.},
  title   = {Peristaltic motion},
  journal = {J. Fluid Mech.},
  year    = {1967},
  volume  = {29},
  number  = {4},
  pages   = {731--743},
}

@Article{chow1972,
	Author = {Chow, J. C. F. and Soda, K.},
	Journal = {Phys. Fluids},
	Pages = {1700--1706},
	Title = {Laminar flow in tubes with constriction},
	Volume = {15},
	Year = {1972}
}

@Article{coenen2021,
	Author = {Coenen, W. and Zhang, X. and Sánchez, A. L.},
	Journal = {J. Fluid Mech.},
	Number = {R2},
	Title = {Lubrication analysis of peristaltic motion in non-axisymmetric annular tubes},
	Volume = {921},
	Year = {2021}
}

@Article{deiber1979,
	Author = {Deiber, J. A. and Schowalter, W. R.},
	Journal = {AIChE Journal},
	Pages = {638--645},
	Title = {Flow through a tube with sinusoidal axial variations in diameter},
	Volume = {25},
	Number = {4},
	Year = {1979}
}

@Article{dodson1971,
	Author = {Dodson, A. G. and Townsend, P. and Walters, K.},
	Journal = {Rheol. Acta},
	Pages = {508--516},
	Title = {On the flow of {Newtonian} and {non-Newtonian} liquids through corrugated pipes},
	Volume = {10},
	Number = {},
	Year = {1971}
}

@Article{lessen1976,
	Author = {Lessen, M. and Huang, P.-S.},
	Journal = {Phys. Fluids},
	Pages = {945--950},
	Title = {Poiseuille flow in a pipe with axially symmetric wavy walls},
	Volume = {19},
	Year = {1976}
}

@Article{jaffrin1971,
	Author = {Jaffrin, M. Y. and Shapiro, A. H.},
	Journal = {Ann Rev Fluid Mech},
	Pages = {13--37},
	Title = {Peristaltic pumping},
	Volume = {3},
	Year = {1971}
}

@Article{ralph1987,
	Author = {Ralph, M. E.},
	Journal = {J.Fluids Engineering},
	Number = {},
	Pages = {255--261},
	Title = {Steady flow structures and pressure drops in wavy-walled tubes},
	Volume = {109},
	Year = {1987},
}

@Article{shapiro1969,
	Author = {Shapiro, A. H. and Jaffrin, M. Y. and Weinberg, S. L.},
	Doi = {10.1017/S0022112069000899},
	Journal = {J Fluid Mech},
	Number = {4},
	Pages = {799--825},
	Publisher = {Cambridge University Press},
	Title = {Peristaltic pumping with long wavelengths at low {R}eynolds number},
	Volume = {37},
	Year = {1969}
}

@Article{sisavath2001,
	Author = {Sisavath, S. and Jing, X. and Zimmerman, R. W.},
	Journal = {Phys. Fluids},
	Pages = {2762--2772},
	Title = {Creeping flow through a pipe of varying radius},
	Volume = {13},
	Year = {2001}
}

@Article{takabatake1988,
	Author = {Takabatake, S. and Ayukawa, K. and Mori, A.},
	Journal = {J Fluid Mech},
	Number = {},
	Pages = {267--283},
	Publisher = {Cambridge University Press},
	Title = {Peristaltic pumping in circular cylindrical tubes: a numerical study of fluid transport
              and its efficiency},
	Volume = {193},
	Year = {1988},
}

@Article{wang2006,
	Author = {Wang, C. Y.},
	Journal = {Phys. Fluids},
	Pages = {078101},
	Title = {Stokes flow through a tube with bumpy wall},
	Volume = {18},
	Year = {2006}
}

\end{document}